\documentclass[10pt,notitlepage,a4paper]{article}
\usepackage{axodraw}

\newcommand{\ie}{{\it i.e.}}
\newcommand{\eg}{{\it e.g.}}

\newcommand{\Pbf}{{\bf P}}

\newcommand{\ieps}{i\epsilon}

\newcommand{\gsim}{\buildrel > \over {_\sim}}

\newcommand{\order}[1]{${\cal O}\left(#1 \right)$}
\newcommand{\morder}[1]{{\cal O}\left(#1 \right)}

\newcommand{\beq}{\begin{equation}}
\newcommand{\eeq}{\end{equation}}
\newcommand{\beqa}{\begin{eqnarray}}
\newcommand{\eeqa}{\end{eqnarray}}
\newcommand{\beqat}{\begin{eqnarray*}}
\newcommand{\eeqat}{\end{eqnarray*}}
\newcommand{\ket}[1]{\vert{#1}\rangle}
\newcommand{\bra}[1]{\langle{#1}\vert}

\newcommand{\PL}[3]{Phys.~Lett.~{\bf {#1}},~{#2}~({#3})}
\newcommand{\NP}[3]{Nucl.~Phys.~{\bf {#1}},~{#2}~({#3})}

\newcommand{\PRD}[3]{Phys.~Rev.~{\bf D{#1}},~{#2}~({#3})}

\newcommand{\PRe}[3]{Phys.~Rep.~{\bf {#1}},~{#2}~({#3})}
\newcommand{\PR}[3]{Phys.~Rev.~{\bf {#1}},~{#2}~({#3})}

\newcommand{\prm}{\textrm{ .}}
\newcommand{\crm}{\textrm{ ,}}

\newcommand{\gz}{\gamma^0}
\newcommand{\go}{\gamma^1}
\newcommand{\gf}{\gamma_5}
\newcommand{\ome}[1]{E_{#1}}

\newcommand{\ud}{\textrm{d}}
\newcommand{\Pri}{\,\textrm{P}}
\newcommand{\Tim}[1]{\mathrm{T}\left\{ {#1} \right\}}

\begin{document}

\begin{flushright}
  hep-ph/0406186\\
  HIP-2004-28/TH
\end{flushright}

\begin{centering}
 {\LARGE Lorentz Contraction of Bound States in 1+1\\ \vspace{5 pt} Dimensional Gauge Theory}

 \vspace{12pt}

 {\large M. J\"arvinen\footnote{Matti.O.Jarvinen@Helsinki.fi}}

 \vspace{7pt}

 {\large Department of Physical Sciences and \\ \vspace{1pt} Helsinki Institute of Physics\\
  \vspace{2pt} POB 64, FIN-00014 University of Helsinki, Finland}

 \vspace{3pt}

\end{centering}

\section*{Abstract}

We consider the Lorentz contraction of a fermion-antifermion bound state in 1+1 dimensional QED.
In 1+1 dimensions the absence of physical, propagating photons allows us to
explicitly solve the weak coupling limit $\alpha\ll m^2$ of the Bethe-Salpeter bound state equation in any Lorentz frame.
In a time-ordered formalism it is seen that all pair production is suppressed in this limit.
The wave function is shown to contract while the mass spectrum is invariant under boosts.

\section{Introduction}

The aim of this paper is to study the boost properties of bound state equal time wave functions in gauge field theories.
We will show explicitly how the fermion-antifermion wave functions Lorentz contract in the case of 1+1 dimensional QED in the weak coupling limit $e^2\ll m^2$. The solution is easier in 1+1 dimensions than in 3+1 dimensions where contributions from propagating gauge
fields must be taken into account.
To our knowledge, the exact Lorentz contraction in gauge field theories has not been demonstrated explicitly before
even in the 1+1 dimensional case.
In \cite{Brodsky} a Lorentz contracting wave function of a two body bound state in (3+1 dimensional) QED is represented as
an approximation valid for small boosts.
The frame dependence of bound state wave functions has also been studied previously in various
models, see for example \cite{Gloeckle,Hoyer}. 
In \cite{Guinea} Lorentz contraction is obtained for a fermion pair interacting via a $\delta$-potential. The frame dependence of
meson wave functions in the Gross-Neveu model is studied in \cite{Schon}.
There has also been other interesting work on the Lorentz covariance of two body equations \cite{Hanson,Artru}.

In the standard instant form quantization of the theory, the wave functions are evaluated with all constituents having the same time
$x^0$. Another possibility is to use the light-front quantization: the theory is quantized at equal light-front time $x^+=(x^0+x^1)/\sqrt{2}$ \cite{Dirac}.
In that case it is natural to study bound state wave functions evaluated at equal $x^+$ instead of $x^0$.

Special relativity leads to Lorentz contraction of classical objects.
According to the Lorentz transformation rules, a moving
object contracts in the direction of motion: the spatial distance between two points of the object is proportional to $1/\gamma$ when measured at equal $x^0$ in the observer's frame. Similarly, the wave functions evaluated at equal $x^0$ of the constituents are expected to Lorentz contract under boosts.
The boost generators $K^i$ do not commute
with the Hamiltonian for translations in $x^0$ and thus the boosts are dynamical. Consequently, the equal $x^0$ bound state wave function of the center-of-mass frame cannot be kinematically
boosted to a general frame. This is why its contraction is non-trivial.

The non-trivial transformation under boosts of wave functions defined at equal $x^0$ is consistent with the Lorentz covariant formulation of bound state equations
(Bethe-Salpeter formalism \cite{Bethe}).
{\it{E.g.}}, a covariant two-fermion wave function for a bound state $\ket{\Pbf}$ with 3-momentum $\Pbf$ is defined by
\beq \label{covwf}
 \chi_{\Pbf}(X,x)_{\alpha\beta}=\bra{\Omega}\Tim{\bar\psi_\beta(X-x/2)\psi_\alpha(X+x/2)}\ket{\Pbf} \prm
\eeq
where $x$ is the relative coordinate.
The Lorentz transformation of this wave function (for a scalar state)
is given by \cite{Brodsky}
\beq \label{strans}
 \chi_{\Pbf'}(X',x')=S(\Lambda)\,\chi_{\Pbf}(X,x)\,S^{-1}(\Lambda)
\eeq
where $x'=\Lambda x$, $P'=\Lambda P$ and $S(\Lambda)$ is the usual spin transformation matrix. For boosts we cannot have both $x^0=0$ and $x'^0=0$. Thus (\ref{strans}) does not
specify a transformation between equal $x^0$ wave functions. However, in \cite{Brodsky} it is suggested that the $x^0$ dependence
of the wave function can be omitted in the weak coupling limit for boosts from the center-of-mass frame to a frame with small momentum $\Pbf$.
We are going to discuss the validity of this approximation here.

The dynamics involved in boosting equal $x^0$ wave functions can be qualitatively seen by considering the hydrogen atom,
which (to a good approximation) is non-relativistic in the rest frame. In this frame the typical scale of the kinetic energy of the electron
is $\alpha^2 m_e$ and its typical momentum
is of order $\alpha m_e$. Thus, by momentum conservation, the momenta of exchanged physical (transverse) photons are (at least) $\alpha m_e$.
Hence $E_\gamma=|{\bf p}_\gamma|=\alpha m_e \gg E_e \approx \alpha^2m_e/2$ so that Fock states with transverse photons may be neglected at lowest order.

In a boosted frame the situation is different.
The electron carries approximately a fraction
$m_e/(m_e+m_p)\equiv a$ of the total center-of-mass momentum $P$. Let the electron momenta parallel and perpendicular to the boost be $aP+p_\parallel$ and $p_\perp$, respectively.
For $aP\gg m_e$, the kinematically boosted electron energy is
\beq
 E = \sqrt{(aP+p_\parallel)^2+p_\perp^2+m_e^2} \approx aP + p_\parallel + \frac{(m_e)^2+p_\perp^2}{2aP} \prm
\eeq
The magnitude of the electron
energy fluctuations is given by the second term $p_\parallel$, which also is the typical photon momentum scale.
Thus in the boosted frame the photon and electron energy fluctuations are of the same order and Fock states with additional photons can be relevant even at lowest order. The wave function at equal $x^0$ of the hydrogen atom in fact has not (to our knowledge) been calculated in a general frame.

We will here study Lorentz contraction in the particularly simple case of 1+1 dimensional QED at weak coupling, $e^2\ll m_e^2$.
In 1+1 dimensions, the time derivatives of the photon field vanish from the Lagrangian
in the gauge $A^1=0$. Thus there are no physical, propagating photons. The photon propagator becomes instantaneous, it only connects fermions
having the same $x^0$.
Our study applies as such to QCD$_2$ which has the same Feynman rules as QED$_2$ up to colour factors in the
$A^{a\,1}=0$ gauge where the gluon self interactions are absent.

Lorentz covariance is nevertheless non-trivial even in 1+1 dimensions. This is illustrated by the fact that the ``relativistic'' Schr\"odinger two body equation with the Hamiltonian
\beq \label{rsham}
 H = \sqrt{(p_1^1)^2+m_1^2}+\sqrt{(p_2^1)^2+m_2^2} + V(|x_1^1-x_2^1|)
\eeq
and a linear (Coulomb) potential is known not to be Lorentz covariant
\cite{Artru}: Its energy eigenvalues do not have the correct dependence on the CM momentum, $E\ne\sqrt{M^2+(P^1)^2}$.
This indicates the importance of including Fock states with additional particle pairs. Here we eliminate their contribution by restricting ourselves to the weak coupling limit of QED, $e^2 \ll m_e^2$. In this limit we show that the eigenvalues of (\ref{rsham}) indeed satisfy $E=\sqrt{M^2+(P^1)^2}$.

\section{Lorentz Contraction in QED$_2$}

We show here that in the weak coupling limit the fermion-antifermion equal $x^0$ wave function in 1+1 dimensional QED (QED$_2$)
Lorentz contracts while the mass spectrum of the equation is invariant in boosts.
The QED$_2$ spectrum is affected by the presence of a background electric field --- the $\theta$ parameter of Ref. \cite{ColemanSchw}.
Since the field strength $F_{01}$ is invariant under boosts we may limit ourselves to the simplest case where the background field vanishes
in all frames. We also take the fermion masses to be equal for notational simplicity.

We use the $A^1=0$ gauge. In this gauge, the photon propagators are infra-red singular. These singularities are handled using the principal value prescription (see, \eg, \cite{tHooft2d}). The propagator is defined to be
\beq \label{proppres}
 D(k^1) \equiv \Pri \frac{i}{(k^1)^2} \equiv \frac 12 \frac{i}{(k^1+\ieps)^2} + \frac 12 \frac{i}{(k^1-\ieps)^2} \prm
\eeq
Then the Fourier transform of the propagator is the linear Coulomb potential
\beq \label{linpot}
 \int \frac{\ud k^1}{2\pi} iD(k^1) e^{ik^1x^1} = -\Pri \int \frac{\ud k^1}{2\pi} \frac{1}{(k^1)^2} e^{ik^1x^1} = \frac 12 |x^1| \prm
\eeq
Note that (\ref{proppres}), (\ref{linpot}) define a principal value integral for a double pole.

\begin{figure}
 \SetScale{0.7}\begin{picture}(300,80)(-33,0)
  \Oval(70,50)(40,30)(0)
  \SetWidth{1.5}
  \DashLine(10,50)(40,50){7}
  \SetWidth{0.5}
  \ArrowLine(130,85)(85,85)
  \ArrowLine(85,15)(130,15)
  \Vertex(130,15){1.5}
  \Vertex(130,85){1.5}
  \Text(107,35)[]{$=$}
  \Oval(230,50)(40,30)(0)
  \SetWidth{1.5}
  \DashLine(170,50)(200,50){7}
  \SetWidth{0.5}
  \ArrowLine(290,85)(245,85)
  \ArrowLine(245,15)(290,15)
  \Boxc(315,50)(50,78)
  \ArrowLine(365,85)(340,85)
  \Oval(375,85)(10,10)(0)
  \Line(385,85)(410,85)
  \ArrowLine(385,15)(410,15)
  \Oval(375,15)(10,10)(0)
  \Line(340,15)(365,15)
  \Vertex(410,15){1.5}
  \Vertex(410,85){1.5}
  \Text(221,35)[]{$K$}
  \Text(263,35)[]{$S$}
  \Text(50,35)[]{$\Psi_{P^1}$}
  \Text(163,35)[]{$\Psi_{P^1}$}
  \Text(0,40)[]{$P$}
  \Text(87,0)[]{$p$}
  \Text(87,71)[]{$P-p$}
  \Text(280,0)[]{$p$}
  \Text(188,0)[]{$k$}
 \end{picture}
 \caption{The Bethe-Salpeter equation. The blobs represent the wave function $\Psi_{P^1}$, $K$ is the interaction kernel and $S$ is the two-particle propagator.}
 \label{waveeq}
\end{figure}
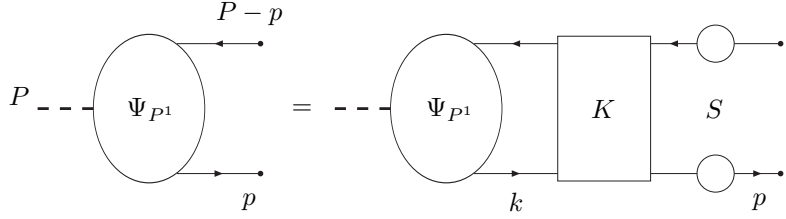

We consider the weak coupling limit of the Bethe-Salpeter bound state equation \cite{Bethe} for QED$_2$ and an arbitrary Lorentz frame.
This straightforward calculation is done more in detail in \cite{Jarv}.
The equation is shown diagrammatically in fig. \ref{waveeq}.
In particular, $S$ is the full two particle propagator including radiative corrections (as in \cite{tHooft2d}). The interaction kernel $K$
contains all possible (irreducible) interaction graphs.
The total two-momentum of the bound state is denoted by $P$.
The wave function $\Psi_{P^1}$ is defined as the coupling of the bound state $\ket{P^1}$ to a fermion pair
\beq
 \Psi_{P^1}(p) = \int \ud^2 x\,e^{ix\cdot p} \bra{\Omega}\Tim{\bar\psi(0)\psi(x)}\ket{P^1}
\eeq
where $\ket\Omega$ is the vacuum of the theory. The equation reads
\beq
 \Psi_{P^1}(p)= S(P,p)\int \frac{\ud^2 k}{(2 \pi)^2}K(P,p,k) \Psi_{P^1}(k) \prm
\eeq
$\Psi_{P^1}$, $S$ and $K$ also have  Dirac indices which were hidden above.

Although the equation (fig. \ref{waveeq}) is for the coupling of the bound state to two particles, there
still is pair production and Fock states with infinitely many particles involved in $K$ and $S$. In order to make this more explicit we now move to a time-ordered formalism
by taking the Fourier transform over $p^0$.
In the gauge $A^1=0$ the photon propagator is independent of energy:
the interaction is instantaneous, connecting fermions at the same instant of time. This makes the time-ordered formalism particularly
simple. Taking the Fourier transform to reach ($x^0,p^1$) space, the photon and fermion propagators become
\beqa
 D(x^0,k^1) &\equiv& \delta(x^0) \Pri \frac{i}{(k^1)^2} = \delta(x^0) D(k^1)\nonumber\\
 S_F(x^0,p^1) &\equiv& \Theta(x^0)\Lambda^+(p^1)e^{-i\,x^0\ome{p^1}} + \Theta(-x^0)\Lambda^-(-p^1)e^{i\,x^0\ome{p^1}}
\eeqa
where the projection operators $\Lambda^\pm$ are defined by
\beq \label{prodef}
 \Lambda^\pm(p^1) \equiv \frac{\pm \gz \ome{p^1}\mp \go p^1 + m}{2 E_{p^1}}
\eeq
and $E_{p^1}\equiv\sqrt{m^2+(p^1)^2}$.
In addition to the usual momentum space Feynman rules, there are now integrations over the time of each vertex. These integrations
give energy denominators.

We wish to find the leading terms of the Bethe-Salpeter equation in the weak coupling limit. To do this we need
to find the correct expansion parameter.
In 1+1 dimensions, the non-relativistic two particle Hamiltonian for a system interacting via a Coulomb force is,
in terms of the relative coordinates and in the center-of-mass frame,
\beq
 H = 
 \frac{1}{2\mu}(p^1)^2 + \frac{e^2}{2}|x^1|
\eeq
where $\mu=m/2$. Equating the (relative) kinetic and potential parts we have $\alpha|x^1| \sim \alpha/p^1 \sim (p^1)^2/m$, hence
\beq
 p^1 \sim \left(\frac{\alpha}{m^2}\right)^{1/3}m\crm\quad V \sim \left(\frac{\alpha}{m^2}\right)^{2/3} m \label{cmscales}\prm
\eeq
Thus in 1+1 dimensions the natural dimensionless expansion parameter is $p^1/m\sim\left(\alpha/m^2\right)^{1/3}$, and
we will calculate to first non-trivial order in this parameter.

In a general Lorentz frame with a total center-of-mass momentum $P$,
we expect contraction effects in addition to (\ref{cmscales}):
\beq  \label{scales}
 p^1 \sim \left(\frac{\alpha}{m^2}\right)^{1/3}m \cdot \frac{\mathcal{E}}{m}\crm\quad V \sim \left(\frac{\alpha}{m^2}\right)^{2/3} m\cdot\frac{m}{\mathcal{E}}
\eeq
where $p^1$ is now the relative, internal momentum and $\mathcal{E}\equiv \sqrt{(2m)^2+(P^1)^2}$.

\subsection{Relevant Diagrams for the Equation in the Weak Coupling Limit}

We will now show that in the Bethe-Salpeter equation we can replace the full propagator $S$ by the free propagator and the kernel $K$ by
single photon exchange in the weak coupling limit in all Lorentz frames.
First we analyze the propagator $S$.
We check explicitly the one loop correction (fig. \ref{radcorfig}).
\begin{figure}
 \SetScale{0.7}\begin{picture}(60,45)(-207,24)
   \Line(50,80)(70,80)
   \ArrowArc(170,50)(104.4,163.3,196.7)
   \DashCArc(-30,50)(104.4,-16.7,16.7){5}
   \Line(70,20)(90,20)
   \Text(40,62)[]{$p^1$}
   \Text(33,33)[]{$k^1$}
   \Text(71,37)[]{$p^1-k^1$}
   \Text(-70,35)[]{$-i\Sigma(x^0,p^1)\quad=\quad-i\delta(x^0)\Sigma(p^1)\quad=$}
  \end{picture}
\caption{One-loop radiative correction to the quark propagator. The $\delta$-function comes from the instantaneuos photon propagator which is denoted by a dashed line.}
\label{radcorfig}
\end{figure}

In the center-of-mass frame the result for $\Sigma(p^1)$ is easy to foresee. In the coordinate space, the instantaneous fermion propagator
vanishes like $\exp(-m|x^1|)$ and the fermion can only propagate a distance $1/m$, a magnitude smaller than the size of the
bound state $x^1\sim 1/p^1 \sim (m^2/\alpha)^{1/3} 1/m$ from (\ref{cmscales}).
Thus the fermion feels the potential of the photon at the distance $1/m$ and
the magnitude of the self-energy correction will be similarly
a magnitude smaller than the binding energy scale $\Delta E \sim V$ from (\ref{cmscales}),
\beq
 \Sigma(p^1)\sim V(x^1\sim 1/m) \sim \alpha/m^2\cdot m \ll \Delta E \sim \left(\alpha/m^2\right)^{2/3}\cdot m \prm
\eeq
$\Sigma(p^1)$ contributes a shift in the fermion energies and hence a shift in the final binding energy, hence the
comparison with $\Delta E$ makes sense.

For a general Lorentz frame it is more convenient to do the exact calculation. Using ($x^0,p^1$) space Feynman rules, we get
\beqa
 -i\Sigma(p^1) &=& -2 \alpha \int \ud k^1 D(p^1-k^1)\gz S_F(0,k^1) \gz \nonumber\\
  &=&-i \alpha \Pri \int \ud k^1 \frac{\go k^1 +m}{(k^1-p^1)^2\sqrt{(k^1)^2+m^2}} \prm
\eeqa
where $\Pri$ denotes the principal value and the instantaneous fermion propagator is defined by\footnote{In fact,
the result for $\Sigma(p^1)$ is not sensitive to how the instantaneous propagator is defined: the discontinous
$\gz$ term vanishes in the $k^1$ integration.}
\beq
 S_F(0,k^1)=\frac 12\left(S_F(\epsilon,k^1)+S_F(-\epsilon,k^1)\right) = \frac{-\go k^1 +m}{2 \ome{k^1}} \prm \label{otdiscdef}
\eeq
The $k^1$ integral can be done using, \eg, Mathematica: 
\beq
 -i\Sigma(p^1) = -2 i\alpha\left(-\frac{\go p^1 + m}{(p^1)^2+m^2} +
 \frac{m(\go m - p^1)}{2((p^1)^2+m^2)^{3/2}}\log\frac{\sqrt{(p^1)^2+m^2}-p^1}{\sqrt{(p^1)^2+m^2}+p^1} \right) \prm
\eeq
In the center-of-mass frame $p^1$ is small w.r.t. the fermion mass, and we can expand in powers of $p^1/m$:
\beq
 -i\Sigma(p^1) = -2 i\alpha\left(-\frac 1m - 2 \go \frac {p^1}{m^2}\right)\left(1+\morder{\frac{(p^1)^2}{m^2}}\right) \prm
\eeq
Wee see that the leading contribution is $\propto \alpha/m$, as expected. The scaling result for any $P^1$ is rather complicated, but the order of the expression is smaller than or equal to $\alpha/\mathcal{E}$ (in a general frame $p^1\sim P^1$).
The result should again be compared with the scale of the potential or the binding energy $\Delta E$ (see (\ref{scales})):
\beq
 \Delta E \sim \left(\frac{\alpha}{m^2}\right)^{2/3} m \cdot\frac{m}{\mathcal{E}} = \left(\frac{m^2}{\alpha}\right)^{1/3}\frac{\alpha}{\mathcal{E}}
\eeq
which is larger than $\Sigma(p^1)$ by a factor of $(m^2/\alpha)^{1/3}$ for all $P^1$. Hence $\Sigma(p^1)$ is negligible.

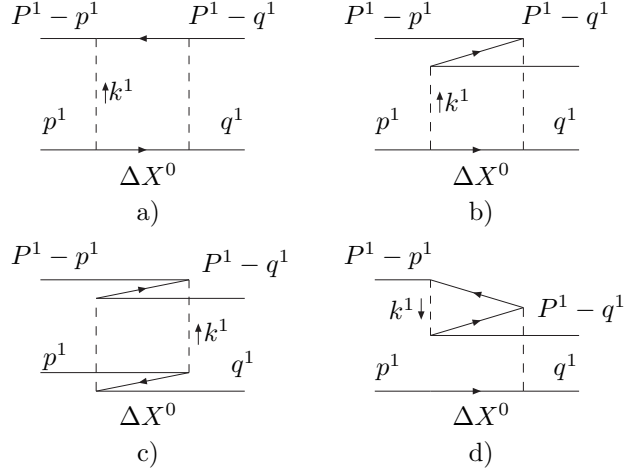
\begin{figure}
\SetScale{0.7}\begin{picture}(140,170)(-30,-100)
 \Line(100,80)(70,80)
 \Line(70,20)(100,20)
 \DashLine(100,20)(100,80){5}
 \ArrowLine(100,20)(150,20)
 \ArrowLine(150,80)(100,80)
 \DashLine(150,20)(150,80){5}
 \Line(180,80)(150,80)
 \Line(150,20)(180,20)
 \Text(55,25)[]{$p^1$}
 \Text(55,65)[]{$P^1-p^1$}
 \Text(122,25)[]{$q^1$}
 \Text(122,65)[]{$P^1-q^1$}
 \Text(90,5)[]{$\Delta X^0$}
 \LongArrow(105,45)(105,55)
 \Text(80,37)[]{$k^1$}

 \Line(330,80)(250,80)
 \Line(250,20)(280,20)
 \DashLine(280,20)(280,65){5}
 \ArrowLine(280,20)(330,20)
 \ArrowLine(280,65)(330,80)
 \DashLine(330,20)(330,80){5}
 \Line(360,65)(280,65)
 \Line(330,20)(360,20)
 \Text(181,25)[]{$p^1$}
 \Text(181,65)[]{$P^1-p^1$}
 \Text(248,25)[]{$q^1$}
 \Text(248,65)[]{$P^1-q^1$}
 \Text(216,5)[]{$\Delta X^0$}
 \LongArrow(285,40)(285,50)
 \Text(207,33)[]{$k^1$}

 \Line(150,-100)(70,-100)
 \Line(70,-50)(150,-50)
 \DashLine(100,-110)(100,-65){5}
 \ArrowLine(150,-100)(100,-110)
 \ArrowLine(100,-60)(150,-50)
 \DashLine(150,-100)(150,-50){5}
 \Line(180,-60)(100,-60)
 \Line(100,-110)(180,-110)
 \Text(55,-64)[]{$p^1$}
 \Text(55,-26)[]{$P^1-p^1$}
 \Text(126,-67)[]{$q^1$}
 \Text(127,-29)[]{$P^1-q^1$}
 \Text(90,-86)[]{$\Delta X^0$}
 \LongArrow(155,-85)(155,-75)
 \Text(116,-54)[]{$k^1$}

 \Line(280,-50)(250,-50)
 \Line(250,-110)(280,-110)
 \DashLine(280,-50)(280,-80){5}
 \ArrowLine(330,-65)(280,-50)
 \ArrowLine(280,-110)(330,-110)
 \ArrowLine(280,-80)(330,-65)
 \DashLine(330,-110)(330,-65){5}
 \Line(360,-80)(280,-80)
 \Line(330,-110)(360,-110)
 \Text(181,-68)[]{$p^1$}
 \Text(181,-26)[]{$P^1-p^1$}
 \Text(248,-68)[]{$q^1$}
 \Text(254,-46)[]{$P^1-q^1$}
 \Text(216,-86)[]{$\Delta X^0$}
 \LongArrow(275,-60)(275,-70)
 \Text(186,-46)[]{$k^1$}

 \Text(90,-10)[]{a)}
 \Text(216,-10)[]{b)}
 \Text(90,-101)[]{c)}
 \Text(216,-101)[]{d)}

\end{picture}

\caption{Examples of ladder diagrams which are present in the full bound state equation. $\Delta X^0>0$ and time flows from left to right.}
\label{ladderg}
\end{figure}

Next we will analyze the structure of $K$ in more detail.
Iterating the bound state equation (fig. \ref{waveeq}) gives the solution as a sum of infinite ladders
of diagrams. When time-ordered, the ladders contain blocks of which a representative set
is shown in fig. \ref{ladderg}. Blocks in fig. \ref{ladderg} a) and c)
consist of single photon exchange diagrams. The diagram in fig. \ref{ladderg} b) is irreducible and comes from a ``crossed'' exhange diagram in $K$ with two photons.
The diagram in fig. \ref{ladderg} d) is a radiative correction to the fermion-photon vertex.
The ``Z graphs'' b), c) and the diagram d) include pair production and are suppressed. The suppression is caused by the $1/\Delta E$ terms arising from the $\Delta X^0$ integration: we have, \eg,
\beqa \label{efrac}
 \mathrm{a)} &\propto& \frac{1}{E-\ome{P^1-p^1+k^1}-\ome{p^1-k^1}} \nonumber\\
 \mathrm{b)} &\propto& \frac{1}{E-\ome{P^1-p^1}-\ome{P^1-q^1-k^1}-\ome{P^1-q^1}-\ome{p^1-k^1}}
\eeqa
where $E$ is the bound state energy. The factors multiplying (\ref{efrac}) are of the same order for both diagrams a) and b).
For diagram a) we have a cancellation
\beq
 E-\ome{P^1-p^1+k^1}-\ome{p^1-k^1} \sim \Delta E \sim \left(\frac{\alpha}{m^2}\right)^{2/3} m \cdot\frac{m}{\mathcal{E}}
\eeq
which does not occur in the case of diagram b), for which the energy difference is \order{\mathcal{E}}. Thus diagram b) is suppressed.
A similar calculation shows the suppression for diagrams c), d) and also for more complicated (irreducible) diagrams with more than two photons, not included in fig. \ref{ladderg}.
The result is that we can neglect all terms but the single photon exchange in $K$ in the weak coupling limit in any Lorentz frame.

\subsection{The Wave Equation in the Weak Coupling Limit}

\begin{figure}
\centering
\SetScale{0.7}\begin{picture}(300,100)(56,0)
 \Oval(90,50)(40,30)(0)
 \SetWidth{1.5}
 \Line(30,50)(60,50)
 \SetWidth{0.5}
 \ArrowLine(150,85)(105,85)
 \ArrowLine(105,15)(150,15)
 \Vertex(150,15){1.5}
 \Vertex(150,85){1.5}
 \LongArrow(56,103)(86,103)
 \Text(50,80)[]{$t$}

 \Text(113,35)[]{=}

 \Oval(240,50)(40,30)(0)
 \SetWidth{1.5}
 \Line(180,50)(210,50)
 \SetWidth{0.5}
 \ArrowLine(300,85)(255,85)
 \ArrowLine(255,15)(300,15)
 \DashLine(300,15)(300,85){5}
 \ArrowLine(300,15)(345,15)
 \ArrowLine(345,85)(300,85)
 \Vertex(345,85){1.5}
 \Vertex(345,15){1.5}
 \Vertex(300,15){1.5}
 \Vertex(300,85){1.5}
 \Text(33,24)[]{$P^1$}
 \Text(96,18)[]{$p^1$}
 \Text(109,0)[]{$X^0$}
 \Text(208,80)[]{$(Y^0>0)$}
 \Text(226,35)[]{$k^1$}
 \Text(251,0)[]{$X^0$}
 \Text(212,0)[]{$X^0-Y^0$}
 \LongArrow(310,60)(310,40)

 \Text(252,35)[]{+}

 \Oval(435,50)(40,30)(0)
 \SetWidth{1.5}
 \Line(375,50)(405,50)
 \SetWidth{0.5}
 \ArrowLine(535,85)(450,85)
 \ArrowLine(450,15)(535,15)
 \DashLine(535,15)(535,85){5}
 \ArrowLine(535,15)(485,30)
 \ArrowLine(485,100)(535,85)
 \Vertex(485,100){1.5}
 \Vertex(485,30){1.5}
 \Vertex(535,15){1.5}
 \Vertex(535,85){1.5}
 \Text(340,80)[]{$(Y^0<0)$}
 \Text(390,35)[]{$k^1$}
 \Text(340,0)[]{$X^0$}
 \Text(374,0)[]{$X^0-Y^0$}
 \Text(65,35)[]{$\varphi_{P^1}$}
 \Text(170,35)[]{$\varphi_{P^1}$}
 \Text(307,35)[]{$\varphi_{P^1}$}
 \LongArrow(545,60)(545,40)\end{picture}
\caption{The time-ordered bound state equation. The blobs denote the equal $x^0$ wave function $\varphi_{P^1}$.}
\label{tobse}
\end{figure}
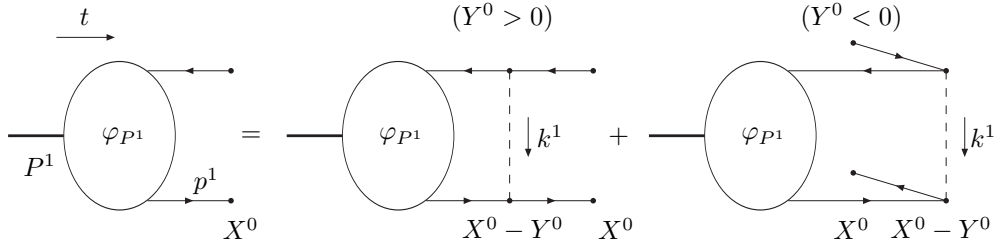

According to the previous section we can use the equation of fig. \ref{waveeq} with free propagators and only single photon exchange in any Lorentz frame.
The result is shown diagrammatically in time ordered form in fig. \ref{tobse}. The analytic expression is
\beqa \label{toeq}
 &\varphi_{P^1}(p^1) = \frac{2\alpha}{E-\ome{p^1}-\ome{P^1-p^1} +\ieps} \Lambda^+(p^1) \gz\, \Pri \int \frac{\ud k^1}{(k^1)^2}\varphi_{P^1}(p^1-k^1)\gz\Lambda^-(P^1-p^1)
 \nonumber\\
 &                +  \frac{2 \alpha}{E+\ome{p^1}+\ome{P^1-p^1} +\ieps} \Lambda^-(p^1)\gz\, \Pri \int \frac{\ud k^1}{(k^1)^2}\varphi_{P^1}(p^1-k^1) \gz\Lambda^+(P^1-p^1)
\eeqa
Here $E$ is the bound state energy and $\varphi$ is the equal time wave function
\beq \label{etwfdef}
 \varphi_{P^1}(p^1)\equiv \int \frac{\ud p^0}{2\pi} \Psi_{P^1}(p)
\eeq
which is a 2x2 matrix in Dirac space in the notation of eq. (\ref{toeq}).

When iterated, the last term on the right hand side of (\ref{toeq}) will generate ``Z graphs'' with pair production (fig. \ref{ladderg} c)).
As shown in the previous section this contribution is suppressed in the weak coupling limit, and the last term can be neglected. This can also
be seen directly from the $1/\Delta E$ terms in equation (\ref{toeq}).

Keeping only the leading term, the equation reads
\beq \label{fteq}
 (E-\ome{p^1}-\ome{P^1-p^1})\varphi_{P^1}(p^1) = 2 \alpha \Lambda^+(p^1)\gz\left[ \Pri \int \frac{\ud k^1}{(k^1)^2}\varphi_{P^1}(p^1-k^1)\right]\gz \Lambda^-(P^1-p^1)
\eeq

From (\ref{fteq}) we see that the wave function $\varphi_{P^1}$ is proportional to the projection matrices $\Lambda^\pm$ which
can be used to further simplify the equation.
When multiplying (\ref{fteq}) with $\Lambda^+(P^1-p^1) \gz$ from the right and $\gz \Lambda^-(p^1)$ from the left we get, using
$\Lambda^+\Lambda^-=0=\Lambda^-\Lambda^+$,
\beqa
  \ome{p}\varphi_{P^1}(p^1)   &=& \left[\gf p^1 + \gz m\right]\varphi_{P^1}(p^1) \nonumber\\ 
  \ome{P-p}\varphi_{P^1}(p^1) &=& \varphi_{P^1}(p^1)\left[ -\gf (P^1-p^1) -\gz m \right]  
\label{deqs}
\eeqa
where $\gf\equiv \gz\go$.
These equations may be used to eliminate the terms proportional to $P^1$, $p^1$ and $m$ in the projection operators  of (\ref{fteq}),
giving on the right hand side
\beq \label{projeq}
 - 2 \alpha \Pri \int \frac{\ud k^1}{(k^1)^2} \frac{\left(\ome{p^1}+\ome{p^1-k^1}\right)+\gf k^1}{2\ome{p^1}} \, \varphi_{P^1}(p^1-k^1)\,
   \frac{\left(\ome{P^1-p^1}+\ome{P^1-p^1+k^1}\right)+\gf k^1}{2\ome{P^1-p^1}} \prm
\eeq
Because of the momentum conservation the photon momentum $k^1$ is of the order of the internal momenta: $k^1 \sim (\alpha/m^2)^{1/3} \mathcal{E} \ll p^1,\; P^1-p^1$.
Hence we can neglect $k^1$ in the numerator of eq. (\ref{projeq}), reducing the projection matrices to unity.
(\ref{fteq}) becomes
\beq \label{seq}
 (E-\ome{p^1}-\ome{P^1-p^1})\varphi_{P^1}(p^1) = -2 \alpha \Pri \int \frac{\ud k^1}{(k^1)^2}\varphi_{P^1}(p^1-k^1) \prm
\eeq
This is the relativistic Schr\"odinger equation, which is not Lorentz covariant for general $\alpha/m^2$ \cite{Artru}. However, further
developing also the energies on left hand side of (\ref{seq}) to first order in $\left(\alpha/m^2\right)^{1/3}$ we have
\beq
 \ome{P^1-p^1}+\ome{p^1} = \mathcal{E} + \frac{(\tilde{p}^1)^2}{2\mu}\frac{(2m)^3}{\mathcal{E}^3}
 +\morder{(\tilde{p}^1)^3} \prm \label{omexp}
\eeq
where $\mu=m/2$, $\mathcal{E}=\sqrt{(2m)^2+(P^1)^2}$ and $\tilde p^1$ is the relative momentum $\tilde p^1 \equiv p^1 - P^1/2$. Fourier transforming into coordinate space we get
\beq \label{sceq}
 (E-\mathcal{E})\varphi_{P^1}(x^1) = \left[-\frac{1}{2\mu}\frac{(2m)^3}{\mathcal{E}^3}\left(\frac{\partial}{\partial x^1}\right)^2 + 2\pi\alpha|x^1|\right] \varphi_{P^1}(x^1) \prm
\eeq
Up to non-leading terms (\ref{sceq}) can be written
\beq \label{schrode}
 \Delta M \varphi_{P^1}(x^1) = \left[-\frac{1}{2\mu}\frac{M^2}{E^2}\left(\frac{\partial}{\partial x^1}\right)^2 + 2\pi\alpha\frac{E}{M}|x^1|\right] \varphi_{P^1}(x^1)
\eeq
where $M$ is the bound state mass and $\Delta M = M - 2 m$. We have thus showed that (\ref{schrode}) is the weak coupling limit of
the Bethe-Salpeter equation (fig. \ref{waveeq}) in an arbitrary Lorentz frame.
Remarkably, we see that
$\varphi_P( x^1/\gamma)$ where $\gamma\equiv E/M$ satisfies an equation which is independent of $P$, and thus the wave function
Lorentz contracts while the spectrum stays invariant. Note that this holds only for a linear potential. Remember that $\varphi_{P^1}$ is a 2x2
matrix in Dirac space although the Dirac structure does not appear in (\ref{schrode}).
The Dirac structure of $\varphi_{P^1}$ is obtained from (\ref{fteq}) (see also section \ref{lfsec} below).

\subsection{Connection to the Covariant Formalism}

In the weak coupling limit the motion of the fermion and the antifermion inside the bound state is slow: $v \sim (\alpha/m^2)^{1/3}$.
Thus it is not essential that the positions of the constituents are measured at exactly the same $x^0$ in the center-of-mass frame. For example, wave functions evaluated at equal $x^0$ and and at equal $x^+$ are expected to coincide. Another consequence, which we prove below, is that the relative time
dependence of the covariant wave function (\ref{covwf}) $\chi_{\Pbf=0}$ can be neglected at leading order in $(\alpha/m^2)^{1/3}$.
This can be used to study more carefully how the transformation rule (\ref{strans}) can be
applied to obtain the Lorentz contraction in the weak coupling limit, \ie, study
the validity of the approximation suggested in \cite{Brodsky}.

We again restrict ourselves to the 1+1 dimensional case with equal fermion masses.
The interaction is instantaneous and backward moving fermions are absent which makes it possible to find
the relative time dependence of the wave function (\ref{covwf}).
The time dependence has the simplest form in ($x^0,p^1$) space using the relative coordinates
\beqa
 \Phi_{P^1}(X^0,x^0,\tilde p^1) &\equiv& \int \ud x^1\,e^{-i\tilde p^1x^1}\, \bra{\Omega}\Tim{\bar\psi(X-x/2)\psi(X+x/2)}\ket{P^1}\Big|_{X^1=0}\nonumber\\
 &=&\int \ud x^1\,e^{-i\tilde p^1x^1}\, \chi_{P^1}(X,x)\Big|_{X^1=0} \prm
\eeqa
where translation invariance allows us to put $X^1=0$.

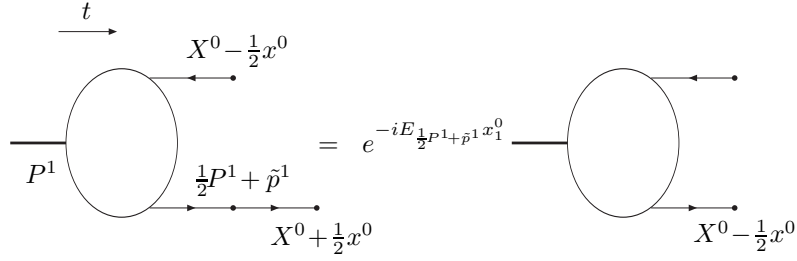
\begin{figure}
\centering
\SetScale{0.7}\begin{picture}(300,85)(6,0)
 \Oval(90,50)(40,30)(0)
 \SetWidth{1.5}
 \Line(30,50)(60,50)
 \SetWidth{0.5}
 \ArrowLine(150,85)(105,85)
 \ArrowLine(150,15)(195,15)
 \ArrowLine(105,15)(150,15)
 \Vertex(150,15){1.5}
 \Vertex(150,85){1.5}
 \Vertex(195,15){1.5}
 \LongArrow(56,110)(86,110)
 \Text(50,85)[]{$t$}

 \Text(173,38)[]{= $\;\; e^{-iE_{\frac 12\!P^1\! + \tilde p^1}x_1^0}$}

 \Oval(360,50)(40,30)(0)
 \SetWidth{1.5}
 \Line(300,50)(330,50)
 \SetWidth{0.5}
 \ArrowLine(420,85)(375,85)
 \ArrowLine(375,15)(420,15)

 \Vertex(420,15){1.5}
 \Vertex(420,85){1.5}
 \Text(33,24)[]{$P^1$}
 \Text(109,21)[]{$\frac 12\!P^1\! + \tilde p^1$}
 \Text(107,70)[]{$X^0\!-\!\frac 12x^0$}
 \Text(139,-1)[]{$X^0\!+\!\frac 12x^0$}

 \Text(299,0)[]{$X^0\!-\!\frac 12x^0$}

 \end{picture}
\caption{The time dependence of the wave function when one of the fermions is shifted in time ($x^0>0$), neglecting the backward moving fermions. The exponential factor arises from the time development of the fermion propagator. The overall time is $X^0$ on the left and $X^0-\frac 12x^0$ on the right.}
\label{tdepfig}
\end{figure}

The overall time dependence of $\Phi$ is given by
\beq \label{tdep}
 \Phi_{P^1} (X^0,x^0,\tilde p^1) = e^{-iEX^0}\Phi_{P^1}(0,x^0,\tilde p^1)
\eeq
where $E=\sqrt{M^2+(P^1)^2}$, $M$ being the bound state mass.
Since the interaction is instantaneous one of the fermions may be shifted in time (see fig. \ref{tdepfig}).
A similar result holds for shifting the antifermion. In terms of $\Phi_{P^1}$ we have
\beq
 \Phi_{P^1}(0,x^0,\tilde p^1) =
  \left\{\begin{array}{rl}
   e^{-iE_{\frac 12\!P^1\! + \tilde p^1}x^0}
   \Phi_{P^1}(-\frac 12x^0,0,\tilde p^1)&\quad\mathrm{when}\;x^0>0\\
   e^{+iE_{\frac 12\!P^1\! - \tilde p^1}x^0}
   \Phi_{P^1}(+\frac 12 x^0,0,\tilde p^1)&\quad\mathrm{when}\;x^0<0
  \end{array}\right.
\eeq
Combining this with (\ref{tdep}) gives the the relative time dependence
\beq \label{rtdep}
 \Phi_{P^1}(0,x^0,\tilde p^1) =
  \left\{\begin{array}{rl}
   e^{-i(E_{\frac 12\!P^1\! + \tilde p^1}-\frac 12E)x^0}
   \Phi_{P^1}(0,0,\tilde p^1)&\quad\mathrm{when}\;x^0>0\\
   e^{-i(\frac 12E-E_{\frac 12\!P^1\! - \tilde p^1})x^0}
   \Phi_{P^1}(0,0,\tilde p^1)&\quad\mathrm{when}\;x^0<0
  \end{array}\right.
\eeq

Let us now apply the 1+1 dimensional version of the transformation (\ref{strans}) to a boost from the center-of-mass frame to a frame with momentum $P^1$ putting $X=0$ and $x'^0=0$:
\beq \label{cmtrans}
 \chi_{P^1}(0,x')\Big|_{x'^0=0} = S(\Lambda)\, \chi_{0}(0,x)\,S^{-1}(\Lambda)
\eeq
Then we have $x^0 = - \sinh\zeta\, x'^1$ with
$\zeta$ denoting the rapidity of the boost. Using the estimates (\ref{scales}) we have
\beq
 x^0 \sim \sinh\zeta\, x'^1 \sim P^1/(m\tilde p'^1) \sim \left(\alpha/m^2\right)^{-1/3} 1/m \cdot P^1/E \prm
\eeq
For the center-of-mass wave function the energy differences appearing in (\ref{rtdep}) are small
\beq
 \frac E2-E_{\pm\tilde p^1} \sim \left(\alpha/m^2\right)^{2/3} m \prm
\eeq
Thus the exponents in (\ref{rtdep}) are $\sim \left(\alpha/m^2\right)^{1/3} P^1/E$ for the center-of-mass wave function
$\Phi_0$ and give a next to leading correction.
This implies that we can drop the relative time dependence of $\chi_0$ in (\ref{cmtrans}).
In terms of the equal $x^0$ wave function
\beq
 \varphi_{P^1}(x^1)=\chi_{P^1}(0,x)\Big|_{x^0=0}
\eeq
we then have
\beq \label{contres}
 \varphi_{P^1}(x^1) = S(\Lambda)\,\varphi_0(\gamma x^1)\, S^{-1}(\Lambda)
\eeq
which is in agreement with the results of the previous section (see also (\ref{speq}) below).

We conclude that the relative time dependence of $\chi_0$ in (\ref{cmtrans}) can be dropped
for all boosts $\Lambda$ at the leading order in $\left(\alpha/m^2\right)^{1/3}$. If next to leading order corrections are included
in the wave functions as in \cite{Brodsky}, (\ref{contres}) is applicable as such for small $P^1/m$.
For large boosts at next to leading order or for higher order corrections the classical contraction result (\ref{contres}) fails.
An analogous result was obtained in the model calculation of Ref. \cite{Hoyer}, where it was seen that the Lorentz contraction differs from the
classical one for $\alpha \gsim m^2$.

\subsection{QED$_2$ in the Infinite Momentum Frame}

\label{lfsec}

It is generally believed (see, \eg, \cite{Pauli}) that the equal $x^0$ description of dynamical processes coincides,
in the infinite momentum frame, with physics on the light-front ($x^+=0$). We may verify this property of our equal $x^0$
bound state wave functions in frames with $P^1\gg 2m$.

The result (\ref{fteq}) implied that $\varphi_{P^1}$ is proportional to the projection matrices $\Lambda^{\pm}$ of (\ref{prodef}),
\beq
 \varphi_{P^1}(\tilde p^1) = \Lambda^+(P^1/2+\tilde p^1)\cdots\Lambda^-(P^1/2-\tilde p^1) \prm
\eeq
in terms of the relative momentum $\tilde p^1 = p^1-P^1/2$.
Taking the contraction result into account, we then have at leading order in $\left(\alpha/m^2\right)^{1/3}$
\beq \label{speq}
 \varphi_{P^1}(\tilde p^1)_{\alpha\beta} = u_\alpha (P^1/2)\bar v_\beta (P^1/2) \phi(\tilde p^1/\gamma)
\eeq
where $\phi$ is a scalar function which is independent of the frame, and  $u,v$ are the Dirac spinors:
\beqa
 u(p^1) &=& \frac{2 E_{p^1}}{\sqrt{E_{p^1}+m}}\Lambda^+(p^1)\left(\begin{array}{c} 1\\0 \end{array}\right) \nonumber\\
 v(p^1) &=& \frac{2 E_{p^1}}{\sqrt{E_{p^1}+m}}\Lambda^-(p^1)\left(\begin{array}{c} 0\\1 \end{array}\right) \prm
\eeqa
In a frame where $P^1\gg 2m$ we get
\beqa \label{slimit}
 u(P^1/2) &=& 
 \sqrt{\frac{P^1}{2}} \left[\left(\begin{array}{c} 1+\frac{m}{P^1}\\1-\frac{m}{P^1} \end{array}\right) + \morder{\frac{m^2}{(P^1)^2}}\right]
 \nonumber\\
 v(P^1/2) &=& 
 \sqrt{\frac{P^1}{2}} \left[\left(\begin{array}{c} 1-\frac{m}{P^1}\\1+\frac{m}{P^1} \end{array}\right) + \morder{\frac{m^2}{(P^1)^2}}\right]
\eeqa
that is, the spinors in (\ref{speq}) approach the form of the light front spinors \cite{Pauli} ($P^1 \approx P^+/\sqrt{2}$ for $P^1\gg 2m$).

The scalar function $\phi(\tilde p^1/\gamma)$ satisfies equation (\ref{seq}). Taking the limit $P^1\rightarrow\infty$ and writing
$E=\sqrt{M^2+(P^1)^2}$ we have
\beq \label{eexp}
 E-E_{p^1}-E_{P^1-p^1} = \frac{M^2}{2P^1}- \frac{m^2}{2 p^1} - \frac{m^2}{2P^1-2p^1} + \morder{\frac{m^4}{(P^1)^3}} \prm
\eeq
if $0<p^1<P^1$. Outside this region, we get an \order{P^1} term in (\ref{eexp}), and inserting this into (\ref{seq}) implies
that the wave function must vanish.
Equation (\ref{seq}) thus reduces to the 't Hooft equation \cite{tHooft2d},
\beq \label{theq}
 M^2\phi(x) = m^2\left(\frac 1x +\frac 1{1-x} \right)\phi(x) - 4 \alpha \Pri \int_0^1\frac{\ud y}{(x-y)^2} \phi(y) \prm
\eeq
where $x \equiv p^1/P^1$. 
This was expected, as all non-planar graphs are suppressed both in the weak coupling limit and in the 't Hooft model.
In the weak coupling limit 't Hooft's model is equivalent to\footnote{Equation (\ref{theq}) does not contain the terms $\propto \alpha\left[1/x+1/(1-x)\right]$ present in 't Hooft's calculation:
these are a next to leading correction in the weak coupling limit.}
light-front QED$_2$.

\section{Discussion}

Hadrons are the true asymptotic states of quantum chromodynamics. Hence
an understanding of scattering processes requires a desciption of
hadrons in arbitrary Lorentz frames. One possibility is to use wave
functions defined at equal light-front time ($x^+=0$) \cite{Pauli}, which
have simple boost properties in the light-front direction. Light-front
wave functions are moreover closely related to cross sections of hard
scattering processes such as deep inelastic scattering. In the bound
state rest frame the choice of a light-front direction obviously
implies a loss of explicit rotational invariance.

Boost properties of bound states wave functions defined at equal $x^0$,
such as Lorentz contraction, are usually discussed in general terms. A
more explicit field theoretic description seems desirable, and could
help in understanding the connection between the quark and parton model
pictures of hadrons. The quark model wave functions defined in the rest
frame should turn into the parton model light-front wave functions when
boosted to infinite momentum. As a first step in this direction we here
considered the boost properties of a non-relativistic hydrogen atom
in 1+1 dimensions.

\subsection*{Acknowledgements}

I would like to thank Paul Hoyer for introducing me to the subject, for several useful discussions and for advice when writing the paper.
I also thank Stanley Brodsky for comments.


\begin{thebibliography}{99}
\bibitem{Brodsky} S. J. Brodsky and J. R. Primack, Annals~Phys.~{\bf 52}~{315}~(1969).
\bibitem{Gloeckle} W. Gl\"ockle and Y. Nogami, \PRD{35}{3840}{1987}.
\bibitem{Hoyer} P. Hoyer, \PL{B172}{101}{1986}.
\bibitem{Guinea} F. Guinea, R. E. Peierls and R. Schieffer, Physica~Scripta~{\bf 33}~282~(1986).
\bibitem{Schon} V. Schon and M. Thies, arXiv:hep-th/0008175.
\bibitem{Hanson} A. J. Hanson, R. D. Peccei and M. K. Prasad, \NP{B121}{477}{1977}.
\bibitem{Artru} X. Artru, \PRD{29}{1279}{1984}.
\bibitem{Dirac} P. A. M. Dirac, Rev.~Mod.~Phys.~{\bf 21},~{392}~(1949).
\bibitem{Bethe} E. E. Salpeter and H. A. Bethe, \PR{84}{1232}{1951}.
\bibitem{ColemanSchw} S. Coleman, Ann.~Phys.~{\bf 101},~{239}~(1976).
\bibitem{tHooft2d} G. 't Hooft, \NP{B75}{461}{1974}.
\bibitem{Jarv} M. J\"arvinen, Master's Thesis, (University of Helsinki, 2004),\\ http://ethesis.helsinki.fi/julkaisut/mat/fysik/pg/jarvinen/quarkant.pdf.
\bibitem{Pauli} S. J. Brodsky, H-C. Pauli and S. S. Pinsky \PRe{301}{299}{1998}.

\end{thebibliography}
\end{document}